\documentclass[conference]{IEEEtran}

\IEEEoverridecommandlockouts
\usepackage{color}
\usepackage{graphicx}
\usepackage{amsmath}
\usepackage{amssymb}
\usepackage{algorithm}
\usepackage{algorithmic}
\usepackage{amsmath}
\usepackage{multirow}
\usepackage{booktabs}
\usepackage{array}
\usepackage{amsthm}
\usepackage{lipsum}
\usepackage{enumerate}
\usepackage{stfloats}
\usepackage{bm}
\usepackage{subfigure}

\newcommand{\be}{\begin{equation}}
\newcommand{\ee}{\end{equation}}
\newcommand{\bea}{\begin{eqnarray}}
\newcommand{\eea}{\end{eqnarray}}
\newcommand{\ba}{\begin{array}}
\newcommand{\ea}{\end{array}}

\newcommand{\non}{\nonumber}

\title{IRS-Enhanced Wideband MU-MISO-OFDM Communication Systems}

\author{ \IEEEauthorblockN{Hongyu Li$^{\dag}$, Rang Liu$^{\dag}$, Ming Li$^{\dag}$, and Qian Liu$^{\ddag}$
\vspace{-0.0 cm} }
\IEEEauthorblockA{$^{\dag}$ School of Information and Communication Engineering   \\  Dalian University of Technology, Dalian, Liaoning 116024, China \\ E-mail: \texttt{\{hongyuli,liurang\}@mail.dlut.edu.cn, mli@dlut.edu.cn} }

\IEEEauthorblockA{$^{\ddag}$ School of Computer Science and Technology \\  Dalian University of Technology, Dalian, Liaoning 116024, China \\ E-mail: \texttt{e-mail: qianliu@dlut.edu.cn}  }}

\begin{document}
\maketitle
\thispagestyle{empty}
\begin{abstract}
Intelligent reflecting surface (IRS) is considered as an enabling technology for future wireless communication systems since it can intelligently change the wireless environment to improve the communication performance.
In this paper, an IRS-enhanced wideband multiuser multi-input single-output orthogonal frequency division multiplexing (MU-MISO-OFDM) system is investigated.
We aim to jointly design the transmit beamformer and the reflection of IRS to maximize the average sum-rate over all subcarriers.
With the aid of the relationship between sum-rate maximization and mean square error (MSE) minimization, an efficient joint beamformer and IRS design algorithm is developed.
Simulation results illustrate that the proposed algorithm can offer significant average sum-rate enhancement, which confirms the effectiveness of the use of the IRS for wideband wireless communication systems.
\end{abstract}


\maketitle

\section{Introduction}

The continuous growth of the number of intelligent devices and the rapid development of emerging services have caused the exponential increase of the demand for wireless network traffic.
This motivates the research on key technologies, such as massive multi-input multi-output (MIMO), ultra-dense network, and the use of millimeter wave (mmWave) bands \cite{Q Wu 2017}, for the fifth-generation (5G) and beyond wireless communications.
However, the above technologies still inevitably face challenges mainly due to the high cost and power consumptions when employing multiple antennas, cells (base stations (BSs)), and/or hardware components (e.g. radio frequency (RF) chains) at mmWave frequencies \cite{C I 2014}. Therefore, it is necessary to find energy-efficient solutions which still provide for high-speed transmissions for future wireless communications.

Recently, the intelligent reflecting surface (IRS), which is a kind of configurable planar surface realized by a large number of hardware-efficient passive reflecting elements (e.g. phase shifters), has been considered as a potential technology for future wireless communication systems \cite{Q Wu 2019}, \cite{J Zhao}.
By adaptively adjusting the elements of the IRS, the propagation environment between the transmitter and the receiver can be dynamically changed. In this way, the channel/beamforming gain can be effectively improved and the communication quality can be enhanced without additional power consumptions.

Many works have been carried out to investigate the IRS designs with focus on power allocation and/or beamformer for both point-to-point single-user (SU-MISO) systems \cite{X Yu 2019}-\cite{Y Han}, and multi-user MISO (MU-MISO) systems \cite{C Huang}, \cite{H Guo} using different metrics (e.g. maximize rate \cite{X Yu 2019}-\cite{Y Han}, \cite{H Guo}, and maximize energy efficiency \cite{C Huang}).
However, the IRS-assisted scenarios mentioned above are restricted to narrowband SU/MU-MISO channels.
When considering more general wideband frequency-selective channels,
the problem will be different and more difficult to be solved since the common IRS should be designed for all subcarriers while the beamformers are given for each subcarrier.
Few work \cite{Y Yang} has studied the IRS-enhanced wideband orthogonal frequency division multiplexing (OFDM) system. The authors in \cite{Y Yang} considered the simplest single-input single-output (SISO) case and provided an iterative algorithm to alternately execute the power allocation and IRS design.
As for more practical wideband multi-user cases, the additional challenge lies in the further complex objective involving inter-user interference.
To the best of our knowledge, IRS-enhanced wideband MU-MISO-OFDM systems have not been investigated in the
literature yet, which motivates our work.

In this paper, we consider a wideband MU-MISO-OFDM system, which is assisted by an IRS with a large number of reflecting elements realized by phase shifters.
We aim to jointly design the beamformer and the reflection of the  IRS whose elements have constant amplitude to achieve maximum average sum-rate over all subcarriers.
Based on the equivalence between sum-rate maximization and mean square error (MSE) minimization, a joint beamformer and IRS design algorithm is proposed.
The performance of the proposed algorithm is validated by extensive simulations, which also confirm the advantages of employing the IRS in wideband wireless communication systems.

\textit{Notations}:
Boldface lower-case and upper-case letters indicate column vectors and matrices, respectively.
$\mathbb{C}$ and $\mathbb{R}^{+}$ denote the set of complex and positive real numbers, respectively.
$(\cdot)^\ast$, $(\cdot)^T$, $(\cdot)^H$, and $(\cdot)^{-1}$ denote the conjugate, transpose, conjugate-transpose operations, and inversion, respectively.
$\mathbb{E} \{ \cdot \}$ represents statistical expectation.
$\Re \{ \cdot \}$ denotes the real part of a complex number.
$\mathbf{I}_L$ indicates an $L \times L$ identity matrix.
$\| \mathbf{A} \|_F$ denotes the Frobenius norm of matrix $\mathbf{A}$.
$\| \mathbf{a} \|_2$ denotes the $\ell_2$ norm of vector $\mathbf{a}$.
$\otimes$ denotes the Kronecker product.
Finally, $\mathbf{A}(i,:)$, $\mathbf{A}(:,j)$, and $\mathbf{A}(i,j)$ denote the $i$-th row, the $j$-th column, and the $(i,j)$-th element of matrix $\mathbf{A}$, respectively. $\mathbf{a}(i)$ denotes the $i$-th element of vector $\mathbf{a}$.

\section{System Model and Problem Formulation}

\begin{figure}
\centering
\includegraphics[height=1.7in]{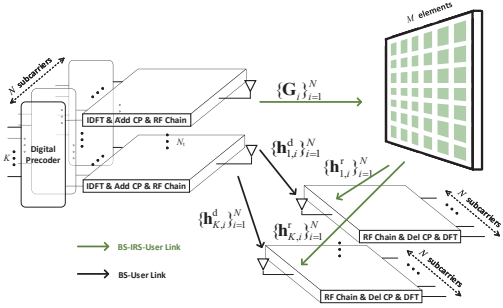}
\caption{An IRS-enhanced wideband MU-MISO-OFDM system.}\label{fig:LIS} \vspace{-0.5 cm}
\end{figure}

We consider a wideband MU-MISO-OFDM system with $N$ subcarriers, as shown in Fig. \ref{fig:LIS}.
The BS employs $N_\mathrm{t}$ antennas to transmit signals to $K$ single-antenna users.
This wireless transmission is assisted by a passive IRS between the BS and users, which employs $M$ phase shifters.
Let $\mathcal{N} = \{1,\ldots,N\}$, $\mathcal{N}_\mathrm{t} = \{1,\ldots,N_\mathrm{t}\}$, $\mathcal{K} = \{1,\ldots,K\}$, and $\mathcal{M} = \{1,\ldots,M\}$ be the set of indices of subcarriers, transmit antennas, users, and elements of the IRS, respectively.
The phase shifters are pointedly adjusted via an IRS controller according to the channel state information (CSI)\footnote{We assume in this paper that the CSI of all channels is known perfectly and instantaneously to the BS. Some recent work also focuses on the channel estimation for IRS-enhanced systems \cite{Taha}, \cite{Zheng}.}.
Next, we will describe the communication process in detail.

\textit{Transmitter:} Let $\mathbf{s}_i = [s_{1,i}, \ldots, s_{K,i}]^T \in \mathbb{C}^{K}$ be the transmit symbols for all users via the $i$-th subcarrier, $\mathbb{E}\{\mathbf{s}_{i}\mathbf{s}_{i}^H\} = \mathbf{I}_{K}$, $\forall i\in \mathcal{N}$. The vector $\mathbf{s}_i$ is first digitally precoded by a precoder matrix $\mathbf{W}_i = [\mathbf{w}_{1,i}, \ldots, \mathbf{w}_{K,i}] \in \mathbb{C}^{N_\mathrm{t} \times K}, \forall i\in \mathcal{N},$ in the frequency domain and then converted to the time domain by the inverse discrete Fourier transform (IDFT), which yields the overall time-domain signal $\widetilde{\mathbf{s}}$ as
\begin{equation}
\widetilde{\mathbf{s}} = (\mathbf{F}^{H}\otimes\mathbf{I}_{N_\mathrm{t}})\mathbf{W}\mathbf{s},
\label{eq:s_timedomain}
\end{equation}
where $\mathbf{F} \in \mathbb{C}^{N \times N},$ is the normalized DFT matrix, $\mathbf{F}(m,n) \triangleq \frac{1}{\sqrt{N}}e^{\frac{-j2\pi (m-1)(n-1)}{N}}, \forall m,n \in \mathcal{N}$.
The overall precoding matrix $\mathbf{W}$ is given by $\mathbf{W} \triangleq \mathrm{diag}(\mathbf{W}_1, \ldots, \mathbf{W}_N)$, and the overall transmit symbol vector can be written as $\mathbf{s} \triangleq [\mathbf{s}_1^T, \ldots, \mathbf{s}_N^T]^T$.
After adding the cyclic prefix (CP) of size $N_\mathrm{cp}$, the signal is up-converted to the RF domain via $N_\mathrm{t}$ RF chains.

\textit{Channel:}
In the considered wideband MU-MISO-OFDM system, the wideband channel from BS to user$_k$ is given by a $D$-tap ($D\le N_\mathrm{cp}$) finite-duration impulse response $\{\widetilde{\mathbf{h}}^{\mathrm{d}}_{k,0}, \ldots, \widetilde{\mathbf{h}}^{\mathrm{d}}_{k,D-1}\}$, where $\widetilde{\mathbf{h}}^{\mathrm{d}}_{k,d} \in \mathbb{C}^{N_\mathrm{t}}$, $d \in \mathcal{D} \triangleq \{0, \ldots, D-1\}$, $\forall k \in \mathcal{K}$, is the impulse response at the $d$-th delay tap.
Similarly, the wideband channel from BS to IRS is given by
$\{\widetilde{\mathbf{G}}_0, \ldots, \widetilde{\mathbf{G}}_{D-1}\}$ involving impulse response  $\widetilde{\mathbf{G}}_d \in \mathbb{C}^{M \times N_\mathrm{t}}$, $\forall d \in\mathcal{D}$.
The wideband channel from IRS to user$_k$ is given by
$\{\widetilde{\mathbf{h}}^{\mathrm{r}}_{k,0}, \ldots, \widetilde{\mathbf{h}}^{\mathrm{r}}_{k,D-1}\}$ with $\widetilde{\mathbf{h}}^{\mathrm{r}}_{k,d} \in \mathbb{C}^{M}, \forall d \in\mathcal{D}, \forall k \in \mathcal{K}$.

\textit{Receiver:}
After propagating through the wideband channels of both the  BS-user link and the BS-IRS-user link, the signal $\widetilde{\mathbf{s}}$ is corrupted by additive white Gaussion noise (AGWN).
After being down-converted to baseband and removing the CP, the time-domain received signal for user$_k$ is given by
\begin{equation}
\widetilde{\mathbf{y}}_k = (\widetilde{\mathbf{H}}^\mathrm{d}_k + \widetilde{\mathbf{H}}^\mathrm{r}_k(\mathbf{I}_{N}\otimes\mathbf{\Phi}) \widetilde{\mathbf{G}})(\mathbf{F}^{H}\otimes\mathbf{I}_{N_\mathrm{t}}) \mathbf{W}\mathbf{s} + \widetilde{\mathbf{n}}_k, \forall k,
\label{eq:y_timedomain}
\end{equation}
where the block cyclic channel matrix $\widetilde{\mathbf{H}}^\mathrm{d}_k \in \mathbb{C}^{N \times NN_\mathrm{t}}$ of the BS-user$_k$ link is defined as
\begin{equation}
\begin{small}
\non
\widetilde{\mathbf{H}}^\mathrm{d}_k = \left[
  \begin{array}{cccc}
     (\widetilde{\mathbf{h}}^{\mathrm{d}}_{k,0})^H &\mathbf{0}_{N_\mathrm{t}}^T &\ldots &(\widetilde{\mathbf{h}}^{\mathrm{d}}_{k,1})^H\\
     \vdots &(\widetilde{\mathbf{h}}^{\mathrm{d}}_{k,0})^H &\vdots &\vdots\\
     (\widetilde{\mathbf{h}}^{\mathrm{d}}_{k,D-1})^H&\vdots &\ddots &(\widetilde{\mathbf{h}}^{\mathrm{d}}_{k,D-1})^H\\
     \mathbf{0}_{N_\mathrm{t}}^T &(\widetilde{\mathbf{h}}^{\mathrm{d}}_{k,D-1})^H &\ddots &\vdots\\
     \vdots &\vdots &\vdots &\mathbf{0}_{N_\mathrm{t}}^T \\
     \mathbf{0}_{N_\mathrm{t}}^T &\mathbf{0}_{N_\mathrm{t}}^T &\ldots &(\widetilde{\mathbf{h}}^{\mathrm{d}}_{k,0})^H\\
  \end{array}
\right], \forall k\in\mathcal{K}.
\end{small}
\end{equation}
Similarly, we define $[\widetilde{\mathbf{G}}^H_0, \ldots, \widetilde{\mathbf{G}}^H_{D-1}, \mathbf{0}_{N_\mathrm{t} \times M}$, $\ldots, \mathbf{0}_{N_\mathrm{t} \times M}]^H$ as the first block column of the block cyclic channel matrix $\widetilde{\mathbf{G}} \in \mathbb{C}^{MN \times NN_\mathrm{t}}$ of the BS-IRS link and $[\widetilde{\mathbf{h}}^{\mathrm{r}}_{k,0}, \ldots, \widetilde{\mathbf{h}}^{\mathrm{r}}_{k,D-1}, \mathbf{0}_{M}, \ldots, \mathbf{0}_{M}]^H$ as the first block column of the block cyclic channel matrix $\widetilde{\mathbf{H}}^\mathrm{r}_k \in \mathbb{C}^{N \times NM}$ of the IRS-user$_k$ link. The phase shift matrix $\mathbf{\Phi}$ of IRS is defined as $\mathbf{\Phi} = \mathrm{diag}(\phi_1, \ldots, \phi_M)$, where each reflecting element has constant amplitude, i.e. $|\phi_m| = 1, \forall m \in\mathcal{M}$, and $\widetilde{\mathbf{n}}_k \in \mathcal{CN}(\mathbf{0}, \sigma^2\mathbf{I}_{N})$ is the AGWN.
After applying DFT, the received signal in the frequency domain can be written as
\begin{equation}
\label{eq:yf_k}
\mathbf{y}_k = \mathbf{F}(\widetilde{\mathbf{H}}^\mathrm{d}_k + \widetilde{\mathbf{H}}^\mathrm{r}_k(\mathbf{I}_{N}\otimes\mathbf{\Phi}) \widetilde{\mathbf{G}})(\mathbf{F}^{H}\otimes\mathbf{I}_{N_\mathrm{t}}) \mathbf{W}\mathbf{s} + \mathbf{n}_k, \forall k,
\end{equation}
where $\mathbf{n}_k \triangleq \mathbf{F}\widetilde{\mathbf{n}}_k, \forall k \in \mathcal{K}$, and the equivalent frequency-domain channel for user$_k$ is given by \cite{SPAWC 2017}
\begin{subequations}
\begin{align}
&\mathbf{F}(\widetilde{\mathbf{H}}^\mathrm{d}_k + \widetilde{\mathbf{H}}^\mathrm{r}_k(\mathbf{I}_{N}\otimes\mathbf{\Phi}) \widetilde{\mathbf{G}})(\mathbf{F}^{H}\otimes\mathbf{I}_{N_\mathrm{t}})\\
\non
\overset{(\textrm{a})}= &\mathbf{F}(\widetilde{\mathbf{H}}^\mathrm{d}_k \mathbf{\Gamma}_1\mathbf{\Gamma}_1^T + \widetilde{\mathbf{H}}^\mathrm{r}_k
\mathbf{\Gamma}_2\mathbf{\Gamma}_2^T (\mathbf{I}_{N}\otimes\mathbf{\Phi}) \mathbf{\Gamma}_2\mathbf{\Gamma}_2^T \widetilde{\mathbf{G}}\mathbf{\Gamma}_1\mathbf{\Gamma}_1^T)\times\\
&~~(\mathbf{F}^{H}\otimes\mathbf{I}_{N_\mathrm{t}}) \mathbf{\Gamma}_1\mathbf{\Gamma}_1^T\\
\non
\overset{(\textrm{b})}= &\mathbf{F}([\widetilde{\mathbf{H}}^\mathrm{d}_{k,1}, \ldots, \widetilde{\mathbf{H}}^\mathrm{d}_{k,N_\mathrm{t}}] + [\widetilde{\mathbf{H}}^\mathrm{r}_{k,1}, \ldots, \widetilde{\mathbf{H}}^\mathrm{r}_{k,M}](\mathbf{\Phi}\otimes\mathbf{I}_{N}) \times\\
&~~\left[\begin{array}{ccc}
     \widetilde{\mathbf{G}}_{1,1} &\ldots &\widetilde{\mathbf{G}}_{1,N_\mathrm{t}}\\
     \vdots &\ddots &\vdots\\
     \widetilde{\mathbf{G}}_{M,1} &\ldots &\widetilde{\mathbf{G}}_{M,N_\mathrm{t}}\\
  \end{array}\right])\times(\mathbf{I}_{N_\mathrm{t}} \otimes\mathbf{F}^{H})\mathbf{\Gamma}_1^T\\
\non
= &[\mathbf{F}\widetilde{\mathbf{H}}^\mathrm{d}_{k,1}\mathbf{F}^H + \mathbf{F}\small{\sum\nolimits_{m=1}^M} \widetilde{\mathbf{H}}^\mathrm{r}_{k,m}\phi_m \widetilde{\mathbf{G}}_{m,1}\mathbf{F}^H, \ldots,\\
&~~\mathbf{F}\widetilde{\mathbf{H}}^\mathrm{d}_{k,N_\mathrm{t}} \mathbf{F}^H + \mathbf{F}\small{\sum\nolimits_{m=1}^M} \widetilde{\mathbf{H}}^\mathrm{r}_{k,m}\phi_m \widetilde{\mathbf{G}}_{m,N_\mathrm{t}}\mathbf{F}^H]\mathbf{\Gamma}_1^T\\
\non
\overset{(\textrm{c})}= &[\bm{\Lambda}^\mathrm{d}_{k,1} +\small{\sum\nolimits_{m=1}^M}\phi_m \bm{\Lambda}^\mathrm{r}_{k,m}\bm{\Xi}_{m,1}, \ldots,\\
&~~\bm{\Lambda}^\mathrm{d}_{k,N_\mathrm{t}} + \small{\sum\nolimits_{m=1}^M} \phi_m \bm{\Lambda}^\mathrm{r}_{k,m}\bm{\Xi}_{m,N_\mathrm{t}}]\mathbf{\Gamma}_1^T\\
\non
\overset{(\textrm{d})}=
&\mathrm{diag}((\mathbf{h}_{k,1}^{\mathrm{d}})^H + (\mathbf{h}_{k,1}^{\mathrm{r}})^H\mathbf{\Phi}\mathbf{G}_1, \ldots, (\mathbf{h}_{k,N}^{\mathrm{d}})^H + \\
\label{eq:hf_f}
&~~(\mathbf{h}_{k,N}^{\mathrm{r}})^H\mathbf{\Phi}\mathbf{G}_N), \forall k,
\end{align}
\end{subequations}
where (a) holds by introducing two column permutation square matrices $\mathbf{\Gamma}_1$ and $\mathbf{\Gamma}_2$ with $\mathbf{\Gamma}_1\mathbf{\Gamma}_1^T = \mathbf{I}_{NN_\mathrm{t}}, \mathbf{\Gamma}_2\mathbf{\Gamma}_2^T = \mathbf{I}_{NM}$, which convert a block cyclic matrix to several cyclic matrices arranged in rows. Specifically, (b) holds by defining cyclic matrices $\widetilde{\mathbf{H}}^\mathrm{d}_{k,n} \in \mathbb{C}^{N\times N}$, $\widetilde{\mathbf{H}}^\mathrm{r}_{k,m} \in \mathbb{C}^{N\times N}$, and $\widetilde{\mathbf{G}}_{m,n} \in \mathbb{C}^{N\times N}$ as
$\widetilde{\mathbf{H}}^\mathrm{d}_{k,n}(:,i) = \widetilde{\mathbf{H}}^\mathrm{d}_{k}(:,n+(i-1)N_\mathrm{t})$, $\widetilde{\mathbf{H}}^\mathrm{r}_{k,m}(:,i) = \widetilde{\mathbf{H}}^\mathrm{r}_{k}(:,m + (i-1)M)$, and $\widetilde{\mathbf{G}}_{m,n}(p,q) = \widetilde{\mathbf{G}}(m + (p-1)M, n + (q-1)N_\mathrm{t})$, $\forall i,p,q \in \mathcal{N},\forall m \in \mathcal{M}, \forall n \in \mathcal{N}_\mathrm{t}, \forall k \in \mathcal{K}$.
Then (c) holds since the DFT matrix can diagonalize the cyclic matrix. Here we define $\bm{\Lambda}^\mathrm{d}_{k,n}, \bm{\Lambda}^\mathrm{r}_{k,m}$, and $\bm{\Xi}_{m,n}$ as the diagonal matrix whose diagonal elements are the corresponding eigenvalues of $\widetilde{\mathbf{H}}^\mathrm{d}_{k,n}$, $\widetilde{\mathbf{H}}^\mathrm{r}_{k,m}$, and $\widetilde{\mathbf{G}}_{m,n}$, respectively.
Finally, (d) holds by defining frequency-domain channels $\mathbf{h}^\mathrm{d}_{k,i} \in \mathbb{C}^{N_\mathrm{t}}$,
$\mathbf{h}^\mathrm{r}_{k,i} \in \mathbb{C}^{M}$, and
$\mathbf{G}_i \in \mathbb{C}^{M \times N_\mathrm{t}}$ as
$\mathbf{h}^\mathrm{d}_{k,i}(n) = (\bm{\Lambda}^\mathrm{d}_{k,n}(i,i))^*$, $\mathbf{h}^\mathrm{r}_{k,i}(m) = (\bm{\Lambda}^\mathrm{r}_{k,m}(i,i))^*$, and $\mathbf{G}_i(m,n) = \bm{\Xi}_{m,n}(i,i)$, $\forall m \in \mathcal{M}, \forall n \in \mathcal{N}_t, \forall k \in \mathcal{K}, \forall i \in \mathcal{N}$.
Substituting (\ref{eq:hf_f}) into (\ref{eq:yf_k}), we can obtain the received signal on the $i$-th subcarrier for user$_k$ as \begin{subequations}
\begin{align}
y_{k,i}=& ((\mathbf{h}_{k,i}^{\mathrm{d}})^H + (\mathbf{h}_{k,i}^{\mathrm{r}})^H\mathbf{\Phi}\mathbf{G}_i) \mathbf{W}_i\mathbf{s}_i + n_{k,i}\\
\non
=& ((\mathbf{h}_{k,i}^{\mathrm{d}})^H + (\mathbf{h}_{k,i}^{\mathrm{r}})^H\mathbf{\Phi}\mathbf{G}_i) \mathbf{w}_{k,i}s_{k,i} + ((\mathbf{h}_{k,i}^{\mathrm{d}})^H+\\
&~~ (\mathbf{h}_{k,i}^{\mathrm{r}})^H\mathbf{\Phi}\mathbf{G}_i) \sum_{p=1,p\ne k}^K\mathbf{w}_{p,i}s_{p,i} + n_{k,i}, \forall k, \forall i,
\end{align}
\end{subequations}
where $n_{k,i}$ denotes the $i$-th element of $\mathbf{n}_{k}$. Then the signal-to-interference-plus-noise ratio (SINR) on the $i$-th subcarrier for user$_k$ is given by
\begin{equation}
\gamma_{k,i} = \frac{|((\mathbf{h}_{k,i}^{\mathrm{d}})^H + (\mathbf{h}_{k,i}^{\mathrm{r}})^H\mathbf{\Phi}\mathbf{G}_i) \mathbf{w}_{k,i}|^2}{\sum_{p\ne k}|((\mathbf{h}_{k,i}^{\mathrm{d}})^H+
(\mathbf{h}_{k,i}^{\mathrm{r}})^H\mathbf{\Phi}\mathbf{G}_i) \mathbf{w}_{p,i}|^2 + \sigma^2}, \forall k, \forall i.
\end{equation}

In this paper, our goal is to jointly design the beamformer $\mathbf{W}$ and the phase shift matrix $\mathbf{\Phi}$ to maximize the average sum-rate for the MU-MISO-OFDM system, subject to the constraints of the phase shift matrix and the transmit power constraint. Therefore, the joint beamformer and IRS design problem can be formulated as
\begin{subequations}
\label{eq:problem0}
\begin{align}
\max_{\mathbf{W}, \mathbf{\Phi}} ~~ &\frac{1}{N}\sum_{i=1}^N\sum_{k=1}^K \log_2(1 + \gamma_{k,i})\\
\label{eq:p0_b}
\mathrm{s.t.} ~~~~ &|\phi_m| = 1, \forall m,\\
\label{eq:p0_c}
&\sum_{i=1}^N\|\mathbf{W}_i\|_F^2 \le P,
\end{align}
\end{subequations}
where $P$ is the total transmit power.

\section{Joint beamformer and IRS design}
\subsection{Problem Reformulation}
Problem (\ref{eq:problem0}) is difficult to solve due to the complex form of the objevtive and the non-convex constraint of the phase shift matrix $\mathbf{\Phi}$. To effectively solve problem (\ref{eq:problem0}), we reformulate the original sum-rate maximization problem as a modified MSE minimization problem \cite{Q Shi 2011}.
Let us first define the modified MSE function for user$_k$ on the $i$-th subcarrier as
\begin{subequations}
\begin{align}
\mathsf{MSE}_{k,i} =
&\mathbb{E}\{(\varpi_{k,i}^\ast y_{k,i} - s_{k,i})(\varpi_{k,i}^\ast y_{k,i} - s_{k,i})^\ast\}\\
\non
=&\sum_{p=1}^K|\varpi_{k,i}^\ast((\mathbf{h}_{k,i}^{\mathrm{d}})^H+
(\mathbf{h}_{k,i}^{\mathrm{r}})^H\mathbf{\Phi}\mathbf{G}_i)\mathbf{w}_{p,i}|^2\\
\non
&~~-2\Re\{\varpi_{k,i}^\ast((\mathbf{h}_{k,i}^{\mathrm{d}})^H+
(\mathbf{h}_{k,i}^{\mathrm{r}})^H\mathbf{\Phi}\mathbf{G}_i)\mathbf{w}_{k,i}\}\\ &~~+ |\varpi_{k,i}|^2\sigma^2+ 1, \forall k, \forall i,
\end{align}
\end{subequations}
where $\varpi_{k,i} \in \mathbb{C}, \forall k \in \mathcal{K}, \forall i \in \mathcal{N},$ is an auxiliary variable.
By introducing the weighting parameter $\rho_{k,i} \in \mathbb{R}^+, \forall k \in \mathcal{K}, \forall i \in \mathcal{N}$, problem (\ref{eq:problem0}) can be equivalently transformed into the following form \cite{Q Shi 2011}:
\begin{subequations}
\label{eq:problem1}
\begin{align}
\label{eq:p1_a}
\max_{\mathbf{W}, \Phi, \bm{\rho, \varpi}} ~~ &\frac{1}{N}\sum_{i=1}^N\sum_{k=1}^K (\log_2(\rho_{k,i}) - \rho_{k,i}\mathsf{MSE}_{k,i} + 1)\\
\mathrm{s.t.} ~~~~ &\textrm{(\ref{eq:p0_b}), (\ref{eq:p0_c})},
\end{align}
\end{subequations}
where $\bm{\rho}$ and $\bm{\varpi}$ denotes the set of variables $\rho_{k,i}$ and $\varpi_{k,i}$, $\forall k \in \mathcal{K}, \forall i \in \mathcal{N}$, respectively.
Now the newly formulated problem (\ref{eq:problem1}) is more tractable than the original problem after removing the complex fractional term (i.e. SINRs) from the $\log(\cdot)$ term.
In particular, problem (\ref{eq:problem1}) is a typical multi-variable-optimization problem, which can be solved using classical block coordinate descent (BCD) iterative algorithms \cite{Bertsekas 1999}.
In the following subsection, we will decompose problem (\ref{eq:problem1}) into four block optimizations and discuss the solution for each block in detail.

\subsection{Block Update}
\subsubsection{Weighting parameter $\bm{\rho}$}
Fixing beamformers $\mathbf{W}_i, \forall i \in \mathcal{N}$, phase shift matrix $\mathbf{\Phi}$, and auxiliary variables $\varpi_{k,i}, \forall k \in \mathcal{K}, \forall i \in \mathcal{N}$, the sub-problem with respect to the weighting parameter $\rho_{k,i}$ is given by
\begin{equation}
\max_{\rho_{k,i}} ~~ \log_2(\rho_{k,i}) - \rho_{k,i}\mathsf{MSE}_{k,i}, \forall k, \forall i,
\label{eq:sub_rho}
\end{equation}
and the optimal solution can be easily obtained by checking the first-order optimality condition of problem (\ref{eq:sub_rho}), i.e.
\begin{equation}
\rho_{k,i}^\star = \mathsf{MSE}_{k,i}^{-1} = 1 + \gamma_{k,i}, \forall k, \forall i,
\label{eq:opt_rho}
\end{equation}

\subsubsection{Auxiliary variable $\bm{\varpi}$}
When the beamformers $\mathbf{W}_i, \forall i \in \mathcal{N}$, phase shift matrix $\mathbf{\Phi}$, and weighting parameters $\rho_{k,i}, \forall k \in \mathcal{K}, \forall i \in \mathcal{N}$, are all fixed, the sub-problem with respect to the auxiliary variable $\varpi_{k,i}$ can be formulated as
\begin{equation}
\min_{\varpi_{k,i}} ~~ \rho_{k,i}\mathsf{MSE}_{k,i}, \forall k, \forall i,
\label{eq:sub_varpi}
\end{equation}
which is a convex unconstrained problem. Thus, problem (\ref{eq:sub_varpi}) can be solved by setting the partial derivative of the objective in (\ref{eq:sub_varpi}) with respect to $\varpi_{k,i}$ to zero, which yields the optimal value of $\varpi_{k,i}$ as
\begin{equation}
\varpi_{k,i}^\star = \frac{((\mathbf{h}_{k,i}^{\mathrm{d}})^H + (\mathbf{h}_{k,i}^{\mathrm{r}})^H\mathbf{\Phi}\mathbf{G}_i) \mathbf{w}_{k,i}}{\sum_{p=1}^K|((\mathbf{h}_{k,i}^{\mathrm{d}})^H+
(\mathbf{h}_{k,i}^{\mathrm{r}})^H\mathbf{\Phi}\mathbf{G}_i) \mathbf{w}_{p,i}|^2 + \sigma^2}, \forall k, \forall i.
\label{eq:opt_varpi}
\end{equation}

\subsubsection{Beamformer $\mathbf{W}$}
With fixed weighting parameters $\rho_{k,i}$, auxiliary variables $\varpi_{k,i}, \forall k \in \mathcal{K}, \forall i \in \mathcal{N}$, and phase shift matrix $\mathbf{\Phi}$, the sub-problem with respect to the beamformer $\mathbf{W}_i, \forall i \in \mathcal{N}$, can be written as
\begin{subequations}
\label{eq:sub_w}
\begin{align}
\non
&\min_{\mathbf{W}}\frac{1}{N}\sum_{i=1}^N\sum_{k=1}^K \rho_{k,i}\Big(\sum_{p=1}^K|\varpi_{k,i}^\ast ((\mathbf{h}_{k,i}^{\mathrm{d}})^H+
(\mathbf{h}_{k,i}^{\mathrm{r}})^H\mathbf{\Phi}\mathbf{G}_i)\times \\
\label{eq:sub_w_a}
&~\mathbf{w}_{p,i}|^2-2\Re\{\varpi_{k,i}^\ast((\mathbf{h}_{k,i}^{\mathrm{d}})^H+
(\mathbf{h}_{k,i}^{\mathrm{r}})^H\mathbf{\Phi}\mathbf{G}_i) \mathbf{w}_{k,i}\}\Big)\\
\label{eq:sub_w_c}
&~~~~\mathrm{s.t.}~~\sum_{i=1}^N\|\mathbf{W}_i\|_F^2 \le P.
\end{align}
\end{subequations}
We define the equivalent channel $\mathbf{h}_{k,i} \triangleq \big(\varpi_{k,i}^\ast((\mathbf{h}_{k,i}^{\mathrm{d}})^H+
(\mathbf{h}_{k,i}^{\mathrm{r}})^H\mathbf{\Phi}\mathbf{G}_i)\big)^H$, $\forall k \in \mathcal{K}, \forall i \in \mathcal{N}$. Then problem (\ref{eq:sub_w}) can be rewritten as
\begin{subequations}
\label{eq:sub_w1}
\begin{align}
\label{eq:sub_w1_b}
&\min_{\mathbf{W}}\frac{1}{N}\sum_{i=1}^N\sum_{k=1}^K \Big(\sum_{p=1}^K\rho_{p,i}|\mathbf{h}_{p,i}^H\mathbf{w}_{k,i}|^2 -2\rho_{k,i}\Re\{\mathbf{h}_{k,i}^H\mathbf{w}_{k,i}\}\Big)\\
\label{eq:sub_w1_c}
&~~~~\mathrm{s.t.}~~\textrm{(\ref{eq:sub_w_c})}.
\end{align}
\end{subequations}
The objective (\ref{eq:sub_w1_b}) groups the terms related to the beamformer $\mathbf{w}_{k,i}$ for user$_k$ at the $i$-th subcarrier together, which motivates us to temporarily ignore the transmit power constraint and separately design the unconstrained beamformer $\mathbf{w}_{k,i}, \forall k \in \mathcal{K}, \forall i \in \mathcal{N}$, by considering the following sub-problem:
\begin{equation}
\label{eq:sub_w1}
\min_{\mathbf{w}_{k,i}} ~\sum_{p=1}^K\rho_{p,i}|\mathbf{h}_{p,i}^H\mathbf{w}_{k,i}|^2 -2\rho_{k,i}\Re\{\mathbf{h}_{k,i}^H\mathbf{w}_{k,i}\}, \forall k, \forall i.
\end{equation}
The optimal unconstrained beamformer can be easily given by
\begin{equation}
\widetilde{\mathbf{w}}_{k,i} = (\sum_{p=1}^K\rho_{p,i}\mathbf{h}_{p,i}\mathbf{h}_{p,i}^H)^{-1} \rho_{k,i}\mathbf{h}_{k,i}, \forall k, \forall i.
\label{eq:opt_uncons_w}
\end{equation}
Finally, we propose to obtain the beamformer that satisfies the total transmit power constraint (\ref{eq:sub_w_c}) using a simple normalization, i.e.
\begin{equation}
\mathbf{w}_{k,i}^\star = \frac{\sqrt{P}\widetilde{\mathbf{w}}_{k,i}} {\sqrt{\sum_{i=1}^N\sum_{k=1}^K\|\widetilde{\mathbf{w}}_{k,i}\|_2^2}}, \forall k, \forall i.
\label{eq:opt_w}
\end{equation}

\subsubsection{Phase shift matrix $\mathbf{\Phi}$}
Given weighting parameters $\rho_{k,i}$, auxiliary variables $\varpi_{k,i}$, and beamfomers $\mathbf{W}_i, \forall i \in \mathcal{N}, \forall k \in \mathcal{K}$, the sub-problem with respect to the phase shift matrix $\mathbf{\Phi}$ can be presented as
\begin{subequations}
\label{eq:sub_phi}
\begin{align}
\non
&\min_{\mathbf{\Phi}}\frac{1}{N}\sum_{i=1}^N\sum_{k=1}^K \rho_{k,i}\Big(\sum_{p=1}^K|\varpi_{k,i}^\ast ((\mathbf{h}_{k,i}^{\mathrm{d}})^H+
(\mathbf{h}_{k,i}^{\mathrm{r}})^H\mathbf{\Phi}\mathbf{G}_i)\times \\
\label{eq:sub_phi_a}
&~~\mathbf{w}_{p,i}|^2-2\Re\{\varpi_{k,i}^\ast((\mathbf{h}_{k,i}^{\mathrm{d}})^H+
(\mathbf{h}_{k,i}^{\mathrm{r}})^H\mathbf{\Phi}\mathbf{G}_i) \mathbf{w}_{k,i}\}\Big)\\
\label{eq:sub_phi_b}
&~~~\;\;\mathrm{s.t.}~~|\phi_m| = 1, \forall m.
\end{align}
\end{subequations}
By defining $\bm{\phi} \triangleq [\phi_1, \ldots, \phi_M]^T$, $\overline{h^\mathrm{d}}_{k,p,i} \triangleq (\mathbf{h}_{k,i}^{\mathrm{d}})^H\mathbf{w}_{p,i}$, and $\mathbf{v}_{k,p,i} \triangleq [(\mathbf{h}_{k,i}^{\mathrm{r}})^H\mathrm{diag} (\mathbf{G}_i\mathbf{w}_{p,i})]^H, \forall k,p\in\mathcal{K}, \forall i \in\mathcal{N}$, problem (\ref{eq:sub_phi}) can be rearranged as
\begin{subequations}
\label{eq:sub_phi1}
\begin{align}
\non
&\min_{\bm{\phi}}\frac{1}{N}\sum_{i=1}^N\sum_{k=1}^K \rho_{k,i}\Big(\sum_{p=1}^K|\varpi_{k,i}^\ast (\overline{h^\mathrm{d}}_{k,p,i} + \mathbf{v}_{k,p,i}^H\bm{\phi})|^2\\
&~~~~-2\Re\{\varpi_{k,i}^\ast(\overline{h^\mathrm{d}}_{k,k,i}+
\mathbf{v}_{k,k,i}^H\bm{\phi})\}\Big)\\
\label{eq:obj_phi_b}
=&\min_{\bm{\phi}}~\bm{\phi}^H\mathbf{A}\bm{\phi}
-2\Re\{\bm{\phi}^H\mathbf{b}\},\\
&~~~\;\;\mathrm{s.t.}~~\textrm{(\ref{eq:sub_phi_b})},
\end{align}
\end{subequations}
where we define
\begin{subequations}
\label{eq:obj_phi1}
\begin{align}
\label{eq:A}
\mathbf{A} &\triangleq \sum_{i=1}^N\sum_{k=1}^K \rho_{k,i}|\varpi_{k,i}|^2\sum_{p=1}^K\mathbf{v}_{k,p,i} \mathbf{v}_{k,p,i}^H,\\
\label{eq:b}
\mathbf{b} &\triangleq \sum_{i=1}^N\sum_{k=1}^K\rho_{k,i}\Big(\varpi_{k,i} \mathbf{v}_{k,k,i}-|\varpi_{k,i}|^2\sum_{p=1}^K\mathbf{v}_{k,p,i} \overline{h^\mathrm{d}}_{k,p,i}\Big).
\end{align}
\end{subequations}

\begin{table*}[t] \caption{Complexity for updating each block.}
\begin{center}
\begin{tabular}{ccccc}
\toprule
  Block & Weighting parameter $\bm{\rho}$ & Auxiliary variable $\bm{\varpi}$ & Beamformer $\mathbf{W}$ & Phase shift matrix $\mathbf{\Phi}$ \\
  \midrule
  Complexity & $\mathcal{O}(NK^2N_\mathrm{t}M^2)$ & $\mathcal{O}(NK(K+1)N_\mathrm{t}M^2)$ & $\mathcal{O}((NK(K+1) + N_\mathrm{t})N_\mathrm{t}^2)$ & $\mathcal{O}(NK^2M^2 + I_1(M-1))$\\
 \bottomrule
\end{tabular}\label{Tab_1}
\end{center} \vspace{-0.2 cm}
\end{table*}

Problem (\ref{eq:sub_phi1}) is still difficult to solve due to the constant magnitude constraint of each phase shift element. To effectively solve this problem, we propose to iteratively design each element of the vector $\bm{\phi}$ until convergence. To facilitate this calculation, we first split the objective (\ref{eq:obj_phi_b}) as
\begin{equation}
\begin{aligned}
&\bm{\phi}^H\mathbf{A}\bm{\phi}-2\Re\{\bm{\phi}^H\mathbf{b}\}\\
=&\sum_{m=1}^M\sum_{n=1}^M\mathbf{A}(m,n)\phi_m^\ast\phi_n - 2\Re\{\sum_{m=1}^M\phi_m^\ast\mathbf{b}(m)\}.\qquad
\end{aligned}
\end{equation}
Then the objective function with respect to the element $\phi_m$ is given by
\begin{subequations}
\begin{align}
\non
f(\phi_m)=&\sum_{n \ne m}(\mathbf{A}(m,n)\phi_m^\ast\phi_n \mathbf{A}(n,m)\phi_n^\ast\phi_m)\\
&\qquad +\mathbf{A}(m,m)|\phi_m|^2-2\Re\{\phi_m^\ast\mathbf{b}(m)\}\quad~~\\
\non
\overset{(\textrm{a})}=&2\Re\Big\{\Big(\sum_{n \ne m}\mathbf{A}(m,n)\phi_n-\mathbf{b}(m)\Big)\phi_m^\ast\Big\}\\
&\qquad+\mathbf{A}(m,m)|\phi_m|^2, \forall m.
\end{align}
\end{subequations}
where (a) holds since $\mathbf{A} = \mathbf{A}^H$. Considering the constant magnitude constraint of each phase shift element (i.e. $|\phi_m|=1, \forall m \in \mathcal{M}$), the sub-problem with respect to $\phi_m$ while fixing other elements can be formulated as
\begin{subequations}
\begin{align}
\max_{\phi_m} ~&\Re\Big\{\Big(\mathbf{b}(m)-\sum_{n \ne m}\mathbf{A}(m,n)\phi_n\Big)\phi_m^\ast\Big\}\\
&\mathrm{s.t.}~~|\phi_m|=1,
\end{align}
\end{subequations}
and the conditionally optimal solution can be determined by
\begin{equation}
\label{eq:opt_phi}
\phi_m^\star =
\frac{\mathbf{b}(m)-\sum_{n \ne m}\mathbf{A}(m,n)\phi_n}{|\mathbf{b}(m)-\sum_{n \ne m}\mathbf{A}(m,n)\phi_n|}, \forall m.
\end{equation}
When low-resolution phase shifters are employed to realize the IRS, the corresponding phase values can be obtained using a simple quantization operation, i.e.
\begin{equation}
\label{eq:opt_phi_qt}
\widetilde{\phi}_m^\star = \exp\left\{j\left[\frac{\angle\{\mathbf{b}(m)-\sum_{n \ne m}\mathbf{A}(m,n)\phi_n\}}{\Delta}\right]\times\Delta\right\}, \forall m,
\end{equation}
where $[\cdot]$ denotes the rounding operation, and $\Delta \triangleq 2\pi/2^b$ is the angle resolution controlled by $b$ bits.

\begin{algorithm}[!t]
\caption{Joint Beamformer and IRS Design}
\label{alg:AA}
\begin{algorithmic}[1]
    \REQUIRE $\mathbf{h}_{k,i}^\mathrm{d},\mathbf{h}_{k,i}^\mathrm{r}, \mathbf{G}_{i}, \forall k \in \mathcal{K}, \forall i \in\mathcal{N}$, $P$, $B$.
    \ENSURE $\mathbf{w}_{k,i}^{\star}, \forall k \in \mathcal{K}, \forall i \in\mathcal{N}, \mathbf{\Phi}^\star$.
        \STATE {Initialize $\mathbf{w}_{k,i}, \forall k \in \mathcal{K},\forall i \in\mathcal{N}, \mathbf{\Phi}$.}
        \WHILE {no convergence of objective (\ref{eq:p1_a})}
            \STATE {Update $\rho_{k,i}, \forall k\in\mathcal{K},\forall i\in\mathcal{N}$ by (\ref{eq:opt_rho}).}
            \STATE {Update $\varpi_{k,i}, \forall k\in\mathcal{K},\forall i\in\mathcal{N}$ by (\ref{eq:opt_varpi}).}
            \STATE {Update $\mathbf{w}_{k,i}, \forall k\in\mathcal{K},\forall i\in\mathcal{N}$ by (\ref{eq:opt_uncons_w}) and (\ref{eq:opt_w}).}
            \STATE {Update $\mathbf{A}$ and $\mathbf{b}$ by (\ref{eq:A}) and (\ref{eq:b}).}
            \WHILE {no convergence of $\mathbf{\Phi}$}
                \FOR {$m=1:M$}
                    \STATE {Update $\phi_m$ by (\ref{eq:opt_phi}) or (\ref{eq:opt_phi_qt}).}
                \ENDFOR
            \ENDWHILE
        \ENDWHILE
        \STATE {Return $\mathbf{w}_{k,i}^{\star}, \forall k\in\mathcal{K},\forall i\in\mathcal{N}, \mathbf{\Phi}^\star$.}
\end{algorithmic}
\end{algorithm}

\subsubsection{Summary}
Having approaches to solve the above four sub-problems with respect to $\rho_{k,i}, \varpi_{k,i}, \mathbf{w}_{k,i}, \forall k \in \mathcal{K},\forall i \in \mathcal{N},$ and $\mathbf{\Phi}$, the overall procedure for the joint beamformer and IRS design is finally straightforward. Given appropriate initial values of $\mathbf{w}_{k,i}, \forall k \in \mathcal{K},\forall i \in \mathcal{N},$ and $\mathbf{\Phi}$, we iteratively update the above four blocks in a pre-specified order until convergence.
The proposed joint beamformer and IRS design algorithm is therefore summarized in Algorithm \ref{alg:AA}.

\subsection{Complexity Analysis}

In this subsection, we provide an analysis of the complexity for the proposed joint beamformer and IRS design algorithm. In each iteration, the complexity for updating each block is summarized in Table 1, where the parameter $I_1$ denotes the number of iterations for updating the phase shift matrix $\mathbf{\Phi}$. Therefore, the total complexity of the proposed algorithm is about $\mathcal{O}(I_2 (NK^2N_\mathrm{t}M^2))$ operations under the assumptions $M \gg N_\mathrm{t}, M \gg K$, and the fact that the method for updating $\mathbf{\Phi}$ can converge within limited iterations. The parameter $I_2$ is the number of iterations for Algorithm \ref{alg:AA}.
Simulation results in the following section further verify the efficiency of the proposed algorithm.

\section{Simulation Results}


In this section, we present simulation results to demonstrate the average sum-rate of the proposed joint beamformer and IRS design. In the considered IRS-enhanced MU-MISO-OFDM system, we assume the number of subcarriers is $N = 64$. The number of taps is set as $D=16$ with half non-zero taps modeled as circularly symmetric complex Gaussian (CSCG) random values.
The CP length is set to be $N_\mathrm{cp} = 16$.
The signal attenuation is set as 30 dB at a reference distance 1 m for all channels. The path loss exponent of the BS-IRS channel, the IRS-user channel, and the BS-user channel is set as 2.8, 2.5, and 3.5, respectively.
The noise power at each user is set as $\sigma^2 = -70$ dBm.
In the following simulation results, we assume the distance between the BS and the IRS is fixed to $d_\mathrm{BI} = 50$ m, the distance between the IRS and users is set as $d_\mathrm{IU} = 3$ m. The distance $d_{\mathrm{BU}_k}$ between the IRS and use$_k$ is randomly selected within the range $d_{\mathrm{BU}_k} \in [d_\mathrm{BI}-d_\mathrm{IU},d_\mathrm{BI}+d_\mathrm{IU}], \forall k\in\mathcal{K}$.
\begin{figure}
\centering
\subfigure[]{
{\label{fig:asr_vs_iter}}
\includegraphics[height=2.5 in]{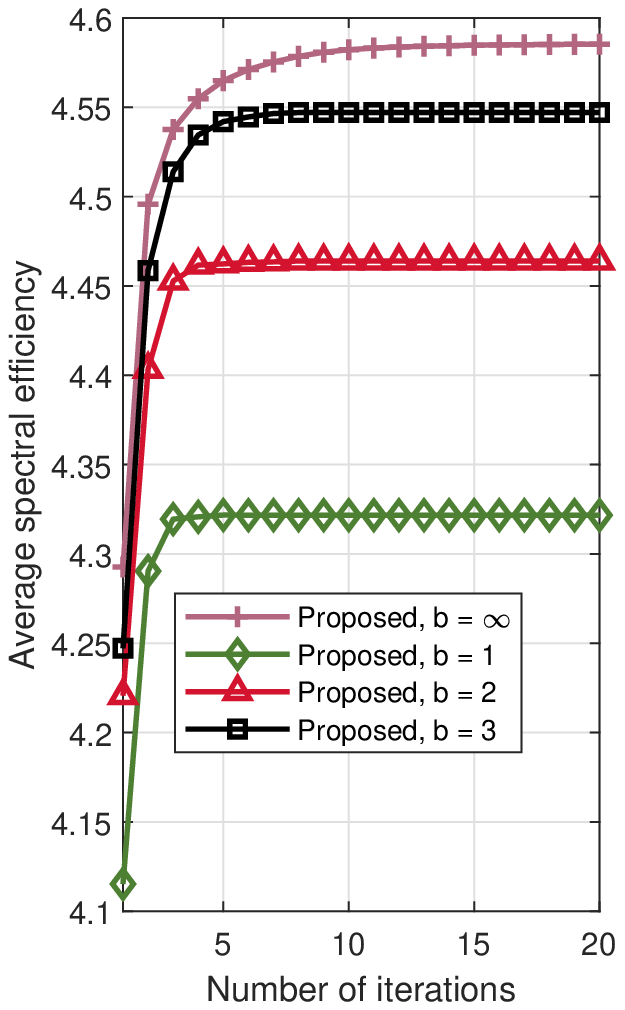}}
\subfigure[]{
{\label{fig:asr_vs_b}}
\includegraphics[height=2.5 in]{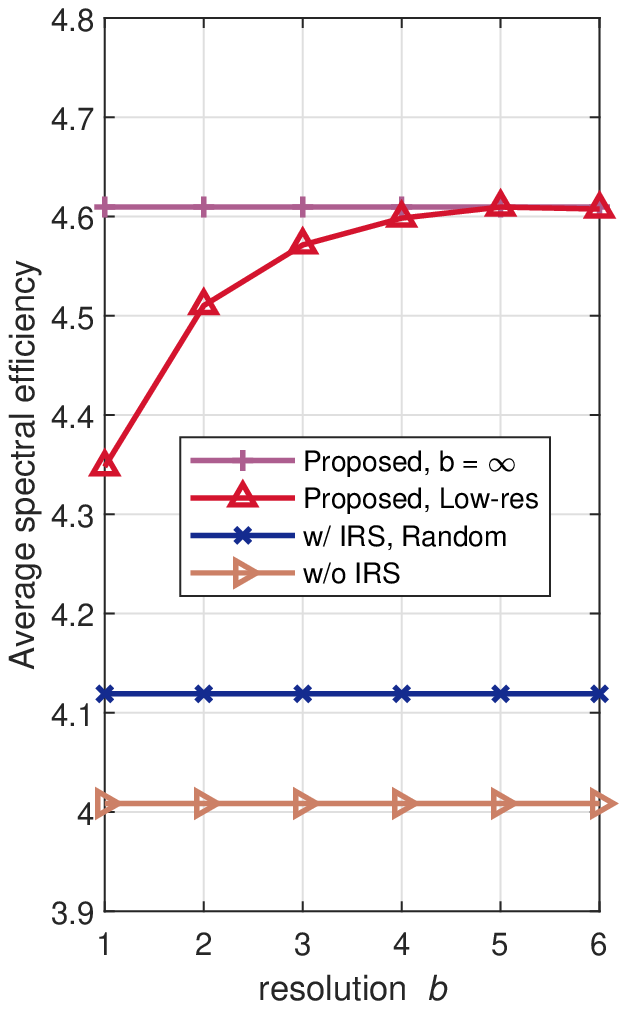}}
\caption{(a) Average spectral efficiency versus the number of iterations; (b) Average spectral efficiency versus the resolution $b$ ($N_\textrm{t} = 8$, $K = 3$, $N = 64$, $M = 64$, $P = 1$W).}\vspace{-0.5 cm}
\end{figure}

We start with presenting the convergence of the proposed joint beamformer and IRS design by plotting the average sum-rate versus the number of iterations in Fig. \ref{fig:asr_vs_iter}.
Simulation results illustrate that the proposed algorithm can converge within 15 iterations when using continuous phase shifters to realize the IRS. For the case of employing low-resolution phase shifters, the proposed algorithm will converge faster within 8 iterations. Combining the complexity analysis in the previous section, the complexity of the proposed algorithm is affordable even with large number of phase shift elements.
Then in Fig. \ref{fig:asr_vs_b}, we plot the average sum-rate as a function of the resolution $b$ (Proposed, Low-res) of each phase shift element. For comparison, we also include the case that each phase shift elemnt of the IRS has random phase and constant amplitude (w/ IRS, Random) as the lower bound. Besides, we plot the average sum-rate achieved by the BS-user link only (w/o IRS). We can observe from
Fig. \ref{fig:asr_vs_b} that there is marginal sum-rate growth beyond $b \ge 4$. Moreover, combining the convergence speed as illustrated in Fig. \ref{fig:asr_vs_iter} and the influence of resolution $b$ as shown in Fig. \ref{fig:asr_vs_b}, using low-resolution phase shifters to realize the IRS is more practical and efficient in realistic systems.

\begin{figure}[!t]
  \includegraphics[width=3.3 in]{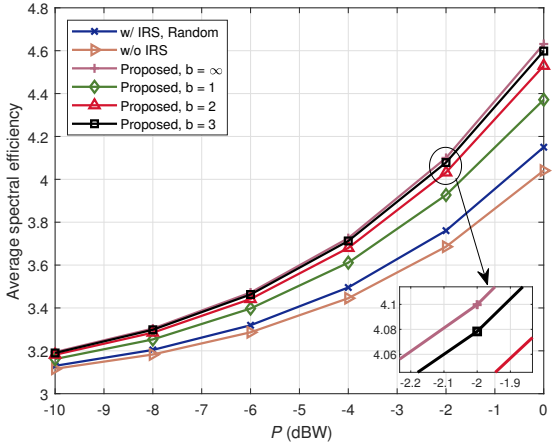}
  \vspace{-0.3 cm}
  \caption{Average sum-rate versus transmit power $P$ ($N_\textrm{t} = 8$, $K = 3$, $N = 64$, $M = 64$).}\label{fig:asr_vs_p}
  \vspace{-0.3 cm}
\end{figure}

\begin{figure}[!t]
  \includegraphics[width=3.3 in]{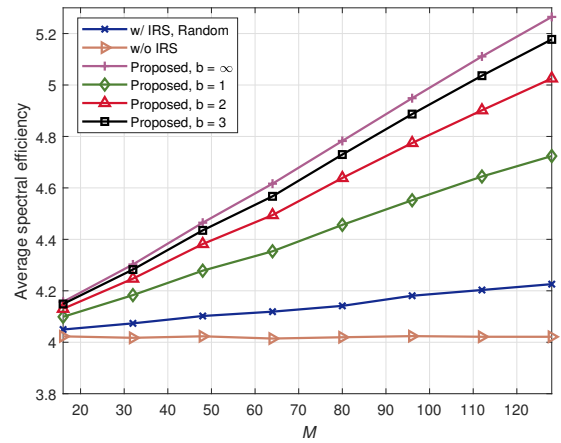}
  \vspace{-0.3 cm}
  \caption{Average sum-rate versus the number of phase shift elements $M$ ($N_\textrm{t} = 8$, $K = 3$, $N = 64$, $P = 1$W).}\label{fig:asr_vs_m}
  \vspace{-0.3 cm}
\end{figure}

Fig. \ref{fig:asr_vs_p} shows the average sum-rate versus the transmit power $P$ with the proposed algorithm for the cases of using continuous and low-resolution (i.e. $b = 1,2,3$-bit) phase shifters. It can be observed that the proposed algorithm can always outperform the ``w/ IRS, Random'' scheme and the ``w/o IRS'' scheme for all transmit power ranges.
When $B = 3$, the proposed algorithm can achieve satisfactory performance close to the case that the IRS is realized by continuous phase shifters, which further confirms the efficiency of employing low-resolution phase shifters.
In Fig. \ref{fig:asr_vs_m}, the average sum-rate versus different numbers of phase shift elements $M$ of the IRS is plotted. A similar conclusion can be drawn from Fig. \ref{fig:asr_vs_m} that the proposed algorithm can always achieve satisfactory performance compared with its competitors, which illustrates the advantages for employing the IRS in wireless communication systems.

\section{Conclusions}
\label{sc:Conclusions}
This paper considered the problem of joint beamformer and IRS design with both continuous and low-resolution PSs to maximize the average sum-rate of a wideband MU-MISO-OFDM system.
We proposed an efficient sub-optimal algorithm with the aid of the equivalence between sum-rate maximization and MSE minimization.
Simulation results demonstrated the advantage of the proposed algorithm, which also revealed the potential of using IRS for wideband wireless communication systems.

\end{document}